# Resonant Combinatorial Frequency Generation Induced by a *PT*-symmetric Periodic Layered Stack

Oksana V. Shramkova, *Senior Member, IEEE*, Giorgos P. Tsironis

*Abstract*—The nonlinear interaction of waves in *PT*-symmetric periodic stacks with an embedded nonlinear anisotropic dielectric layer illuminated by plane waves of two tones is examined. The three-wave interaction technique is applied to study the nonlinear processes. It is shown that the intensity of the three-wave mixing process can be significantly enhanced in resonant cavities based on *PT*-symmetric periodic structures, especially as the pumping wave frequency is near the coherent perfect absorber-lasing resonances. The main mechanisms and properties of the combinatorial frequency generation and emission from the stacks are illustrated by the simulation results and the effect of the layer arrangement in *PT*-symmetric walls of resonator on the stack nonlinear response is discussed. The enhanced efficiency of the frequency conversion at Wolf-Bragg resonances is demonstrated. It has been shown that Wolf–Bragg resonances of very high orders may lead to the global maxima and nulls of the scattered field. The analysis of the effect of losses in nonlinear dielectric layer on the combinatorial frequency generation efficiency has shown that the rate of losses may amplify the intensity of the frequency mixing process.

*Index Terms*—Combinatorial frequency generation, nonlinear dielectric, periodic structure, *PT*-symmetry, three-wave mixing

## I. Introduction

NEW electromagnetic materials and structures that can efficiently control the signals in the millimeter, terahertz and optical ranges, are increasingly sought for the development of novel functional devices. The nonlinear media attract particular attention. Although the nonlinear wave interactions in solids are normally weak, they exhibit unique properties essential for diverse practical applications. For example, it has been shown that metamaterials and photonic crystals with nonlinear inclusions can be instrumental for increasing the efficiency of frequency conversion and harmonic generation in THz and optical ranges [1]-[3]. Since the majority of the currently available metamaterials are fabricated as stacked layers, it is expedient to investigate their nonlinear characteristics in planar layered structures. The recent works in this area have been primarily concerned with the second and third harmonic generation in such type of artificial structures [3]-[8]. Conversely, processes of nonlinear scattering and frequency mixing in finite layered structures illuminated by two or more plane waves of different tones incident at dissimilar angles and the mechanisms of the resonance frequency conversion and combinatorial frequency generation (CFG) still remain scarcely explored and poorly understood. The interest to this topic is connected with the fact that the mixing process in the nonlinear artificial structures can be controlled not by only tailoring the coherence and localisation of the interacting waves but also by altering independently the intensity and phasing of individual pump waves [9]. In earlier studies, it was demonstrated that CFG by two pump waves of frequencies $\omega_1$ and $\omega_2$, incident at dissimilar angles, offers additional degrees of freedom in controlling the frequency mixing process and spectrum [10]-[11].

Parity-time (*PT*)-symmetric systems attract increasing interest fuelled by their applications at the optical range. The fundamental studies of Bender and Boettcher [12] have indicated that even non-Hermitian Hamiltonians can exhibit entirely real spectra as long as they respect the conditions of *PT*-symmetry. In the context of optics, *PT*-symmetric systems have demonstrated several exotic features, including unidirectional invisibility [13], coherent perfect absorption [14], nonreciprocity of light propagation [15]-[16], beam refraction [15] and various extraordinary nonlinear effects [17]. The optical medium, consisting of a uniform index grating with two homogeneous and symmetric gain and loss regions, that realizes a *PT*-symmetric coherent perfect absorber (CPA) laser was introduced in [18]. It was demonstrated that such systems can behave simultaneously as a laser oscillator, emitting outgoing coherent waves, and as a CPA, absorbing incoming coherent waves.

In the present work, the problem of *PT*-symmetric periodic stacks with an embedded nonlinear anisotropic dielectric layer illuminated by plane waves of two tones is revisited, and the properties of reflected and refracted waves of the combinatorial frequencies are examined. The problem statement and the solution of the respective boundary value problem obtained by the three-wave interaction method [19] are outlined in Section 2. The features of CFG by the nonlinear resonant structures based on *PT*-symmetric periodic stacks of the layers, are illustrated by the examples of

This work was partially supported by the European Union Seventh Framework Program (FP7-REGPOT-2012-2013-1) under grant agreement No. 316165.

O. V. Shramkova is with Crete Center for Quantum Complexity and Nanotechnology, Department of Physics, University of Crete, P.O. Box 2208, 71003 Heraklion, Greece (e-mail: oksana@physics.uoc.gr).

G. P. Tsironis is with Crete Center for Quantum Complexity and Nanotechnology, Department of Physics, University of Crete, P.O. Box 2208, 71003 Heraklion, Greece; Institute of Electronic Structure and Laser, Foundation for Research and Technology Hellas, P.O. Box 1527, 71110, Heraklion, Greece; National University of Science and Technology MISiS, Leninsky prosp. 4, Moscow, 119049, Russia (e-mail: gts@physics.uoc.gr).

numerical simulations and discussed in Section 3. The main findings are summarised in Conclusion.

## II. Problem Statement And Main Equations

Let us consider two plane waves of frequencies $\omega_1$ and $\omega_2$ incident at angles $\Theta_{i1}$ and $\Theta_{i2}$, respectively, on a periodic stack composed of the dielectric layers of two types with complex-conjugate dielectric permittivities $\varepsilon = \varepsilon' - i\varepsilon''$ and $\varepsilon^* = \varepsilon' + i\varepsilon''$ ($\varepsilon'$ and $\varepsilon''$ are positive) corresponding to balanced gain and loss regions and identical thickness $d_m$. The structure has a nonlinear dielectric region of length $d_c$ in the centre as shown in Fig. 1. The nonlinear layer has 6mm class of anisotropy and is characterised by the tensors of linear dielectric permittivity $\hat{\varepsilon}_c = (\varepsilon_{xx}, \varepsilon_{xx}, \varepsilon_{zz})$ and a second-order nonlinear susceptibility tensor $\hat{\chi}$. We assume that nonlinearity is weak and the structure is isotropic in the *x-y* plane. Therefore the incident waves of the TE and TM polarisations with the fields independent of the *y*-coordinate ($\partial/\partial y = 0$) can be treated separately. Only the case of TM-polarization is discussed here, whilst the TE waves, unaffected by the anisotropy of $\hat{\chi}$, are analysed similarly. Such a structure can be considered as an optical resonator with a cavity built up by *PT*-symmetric periodic stacks. The stack of total thickness $L=2L_{PT}+d_c$, where $L_{PT}=2Nd_m$, $N$ is the number of unit cells on each side away from the cavity, and is surrounded by linear homogeneous medium with dielectric permittivity $\varepsilon_r$ at $z \leq -L/2$ and $z \geq L/2$.

Making use of the assumption of weak nonlinearity, the harmonic balance method is used next to linearise the system of differential equations. In the non-depleting approximation, this gives the homogeneous differential equations for the pump wave frequencies and non-homogeneous differential equations for the fields of combinatorial frequencies $\omega_3$. The solution of nonhomogeneous wave equation for $H_y$ at frequency $\omega_3$ composed of the particular and general solutions can be cast in the form

$$H_{y3}^L = \left(A^+ e^{ik_z^{(3)}z} + A^- e^{-ik_z^{(3)}z} + N_1^+ e^{ik_z^+ z} + \right.$$
$$\left. + N_2^+ e^{-ik_z^+ z} + N_1^- e^{ik_z^- z} + N_2^- e^{-ik_z^- z}\right)e^{-i\omega_3 t + ik_{x3}x}, \quad (1)$$

where $k_z^\pm = k_z^{(1)} \pm k_z^{(2)}$, $k_z^{(s)} = \sqrt{\varepsilon_{xx}\left(k_s^2 - \dfrac{k_{xs}^2}{\varepsilon_{zz}}\right)}$, $k_s = \omega_s/c$, $s = 1,2,3$, $c$ is the speed of light. The wavenumber $k_{x3} = k_3\sqrt{\varepsilon_r}\sin\Theta_3$ is determined by the requirement of phase synchronism in the three-wave mixing process [19]

$$k_{x3} = k_{x1} + k_{x2}, \quad (2)$$

where $k_{x1,x2} = k_{1,2}\sqrt{\varepsilon_r}\sin\Theta_{i1,i2}$. The angle $\Theta_3$ defines the direction of the combinatorial frequency emission from the stack. Coefficients $A^\pm$ are amplitudes of the waves of frequency $\omega_3$ generated outside the layer and refracted into it, $N_{1,2}^\pm$ are amplitudes of the waves of frequency $\omega_3$ generated inside the layer and these coefficients are expressed in terms of the field magnitudes in the layer at frequencies $\omega_1$ and $\omega_2$.

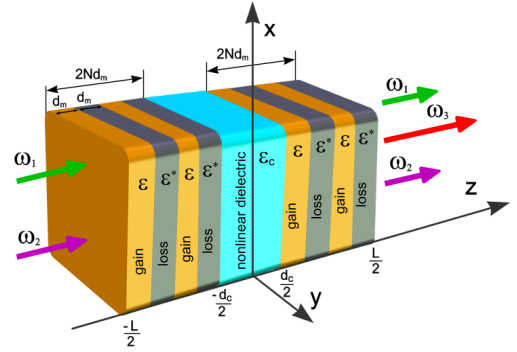

Fig. 1. Schematic of the problem geometry.

In order to examine the three-wave mixing process in the structures, it is necessary to generalise the transfer matrix method (TMM) based analysis for the case of two obliquely incident pump waves [10]. Enforcing the periodic boundary conditions through the whole periodic structure, the fields at the stack outer interfaces can be presented in the form

$$\begin{pmatrix} H_y(\omega_3, -L/2) \\ E_x(\omega_3, -L/2) \end{pmatrix} = \hat{M}(\omega_3)\begin{pmatrix} H_y(\omega_3, L/2) \\ E_x(\omega_3, L/2) \end{pmatrix} + \hat{m}_c(\omega_3)\begin{pmatrix} \tau \\ \xi \end{pmatrix}. \quad (3)$$

Here $\hat{M} = \hat{m}^N \hat{m}_c \hat{m}^N$ is the transfer matrix of the whole stack, $\hat{m}$ is the transfer matrix of the *PT*-symmetric stack unit cell, $\hat{m}_c$ is the transfer matrix of the anisotropic dielectric layer, $\tau$ and $\xi$ contain the terms proportional to coefficients $N_{1,2}^\pm$ for the nonlinear layer [10].

The amplitudes $F_{r,t}$ of the waves scattered from the layer into the surrounding linear media ($F_r$ at $z < -L/2$ and $F_t$ at $z > L/2$) at the combinatorial frequency $\omega_3$ are determined by enforcing the boundary conditions of $H_y(\omega_3)$ and $E_x(\omega_3)$ continuity at the structure interfaces. Finally, we obtain the closed form solution for $F_{r,t}$ is as follows:

$$F_r = \left(M_{21}(\omega_3) + M_{22}(\omega_3)\dfrac{k_{zr}^{(3)}}{k_3\varepsilon_r}\right)\lambda_1 -$$
$$-\left(M_{11}(\omega_3) + M_{12}(\omega_3)\dfrac{k_{zr}^{(3)}}{k_3\varepsilon_r}\right)\lambda_2$$

$$F_t = -\dfrac{k_{zr}^{(3)}}{k_3\varepsilon_r}\lambda_1 - \lambda_2, \quad (4)$$

$$\lambda_r = \frac{m_{cr1}\tau + m_{cr2}\xi}{\Delta}, \quad r = 1, 2;$$

$$\Delta = M_{11}(\omega_3) - \frac{k_{zr}^{(3)}}{k_3} M_{12}(\omega_3) - \frac{k_3}{k_{zr}^{(3)}} M_{21}(\omega_3) + M_{22}(\omega_3).$$

where $k_{zr}^{(3)} = \sqrt{k_3^2 \varepsilon_r - k_{x3}^2}$ is the longitudinal wave number of the wave at frequency $\omega_3$ in the surrounding homogeneous media.

The magnitudes $|F_{r,t}|$ of the field emitted from the layer into the surrounding dielectric media at the combinatorial frequency $\omega_3$ vary with the thickness of nonlinear layer and total thickness of the stack. Unlike the linear dielectric layer, the $|F_{r,t}|$ dependence on the thicknesses of the layers becomes rather intricate in the distributed three-wave mixing processes. In the meantime, Wolf-Bragg resonances in the layer of thickness $L$ equal an integer number of half-waves at the frequency $\omega_3$, can provide qualitative insight in the properties of scattered fields. Applying the resonance condition $k_z^{(3)} d_c = q\pi, q = 1, 2,...$ to eq. (4) the expressions for $F_{r,t}$ can be represented in the following form

$$F_r = \frac{1}{\Delta} \sum_{n=\pm} \left[ (-1)^q \left( M_{21}(\omega_3) + M_{22}(\omega_3) \frac{k_{zr}^{(3)}}{k_3 \varepsilon_r} - \frac{k_z^n}{\varepsilon_{xx}} \left( M_{11}(\omega_3) + M_{12}(\omega_3) \frac{k_{zr}^{(3)}}{k_3 \varepsilon_r} \right) \right) N_1^n - e^{-ik_z^n d_c} \left( M_{21}(\omega_3) + M_{22}(\omega_3) \frac{k_{zr}^{(3)}}{k_3 \varepsilon_r} + \frac{k_z^n}{\varepsilon_{xx}} \left( M_{11}(\omega_3) + M_{12}(\omega_3) \frac{k_{zr}^{(3)}}{k_3 \varepsilon_r} \right) \right) N_2^n \right] \left( 1 - (-1)^q e^{ik_z^n d_c} \right),$$

$$F_t = -\frac{1}{\Delta} \frac{k_{zr}^{(3)}}{k_3 \varepsilon_r} \sum_{n=\pm} \left[ (-1)^q \left( 1 + \frac{\varepsilon_r k_z^n}{\varepsilon_{xx} k_{zr}^{(3)}} \right) N_1^n - e^{-ik_z^n d_c} \left( 1 - \frac{\varepsilon_r k_z^n}{\varepsilon_{xx} k_{zr}^{(3)}} \right) N_2^n \right] \left( 1 - (-1)^q e^{ik_z^n d_c} \right), \quad (5)$$

where $\widehat{M} = \widehat{m}^{2N}$. Inspection of eq. (5) suggests that $|F_{r,t}|$ have nulls and absolute maxima when the following relations are satisfied simultaneously

$$q\pi - k_{zL}^{\pm} d_c = m^{\pm} \pi, \qquad m^{\pm} = 1, 2,... \quad (6)$$

where $m^+$ and $m^-$ have the same parity (nulls at even $m^{\pm}$ and maxima at odd $m^{\pm}$). Since the nonlinear layer acts as a resonant source of the waves emitted into surrounding homogeneous medium at frequency $\omega_3$, both $F_r$ and $F_t$ reach their extremes concurrently.

New feature of a standing wave in the layer is observed at frequency $\omega_3$ when the $z$-components of wave vectors of pump waves in the layer are equal, i.e. $k_z^{(1)} = k_z^{(2)}$. In this case $k_z^- = 0$ and $H_y$ and $E_z$ fields contain the uniform component $(N_1^- + N_2^-)$ in eq.(1), which does not vary across the layer thickness. The latter condition $k_z^{(1)} = k_z^{(2)}$ is satisfied only at the particular relation between frequencies $\omega_{1,2}$ and incidence angles $\Theta_{i1, i2}$

$$\frac{\omega_1}{\omega_2} = \sqrt{\frac{\varepsilon_{zz} - \sin^2 \Theta_{i2}}{\varepsilon_{zz} - \sin^2 \Theta_{i1}}} \quad (7)$$

and the resulting combinatorial frequency $\omega_3$ cannot coincide with the nulls and absolute maxima of the Wolf-Bragg resonances. The results of the parametric analysis presented in the next section further illustrate the properties of the TM waves of combinatorial frequencies $\omega_3$, scattered from the *PT*-symmetric stack with nonlinear anisotropic dielectric layer.

III. PARAMETRIC ANALYSIS AND DISCUSSION

The analytical expressions for $|F_{r,t}|$ provide qualitative insight in the composition and properties of the scattered fields of the combinatorial frequency $\omega_3 = \omega_1 + \omega_2$ and enable analysis of the effects of the structure parameters, incidence angles and frequencies of pump waves on the nonlinear scattering by the resonant structures based on *PT*-symmetric periodic structures. The features of the CFG and emission by optical resonator built up by *PT*-symmetric periodic stacks with an embedded nonlinear anisotropic dielectric layer are illustrated by simulations and are discussed in this section. The *PT*-symmetric stacks with periodic sequences of layers are comprised of layers of two types with $d=100$ μm and $\varepsilon' = 2.0, \varepsilon'' = 0.1$. The following parameters of the embedded nonlinear layer have been used in the simulations: $\varepsilon_{xx} = 5.382$, $\varepsilon_{zz} = 5.457$, $\chi_{xxz} = 2.1 \times 10^{-7}$ m/V, $\chi_{zxx} = 1.92 \times 10^{-7}$ m/V, $\chi_{zzz} = 3.78 \times 10^{-7}$ m/V [20]. The whole stack is surrounded by air with permittivity $\varepsilon_r = 1$.

To gain insight in the properties of CFG, it is necessary first to examine the linear properties of the stack. In particular, we should focus on the phenomenon of spontaneous symmetry breaking.

*A. PT-Symmetry Breaking*

The earlier studies of *PT*-symmetric systems reveal the existence of spontaneous symmetry breaking transition to a phase with a complex eigenspectrum beyond a critical threshold of gain/loss level [21]. It was demonstrated that the electromagnetic scattering matrix (*S*-matrix) $\widehat{S} = \begin{pmatrix} R^{(L)} & T \\ T & R^{(R)} \end{pmatrix}$, where $R^{(L,R)}$ are stack reflection coefficients for wave incident from the left and right, $T$ is the transmission coefficient (it is the same for left and right incidence), measures the breaking of *PT*-symmetry. In the *PT*-symmetric phase the eigenvalues of the *S*-matrix $\lambda_{1,2}$ are unimodular and $|\lambda_1| = |\lambda_2| = 1$, whereas in *PT*-symmetry breaking points both eigenvalues meet and bifurcate, and the broken phases correspond to $|\lambda_1| = 1/|\lambda_2| > 1$.

The frequency dependences of the *S*-matrix eigenvalues for an "ideal" *PT*-symmetric stack without a passive region for wave incident at $\Theta_i = 10°$ is displayed in Fig.2. The exceptional points for which one eigenvalue of the scattering matrix tends to infinity while the second goes to zero correspond to the CPA laser points. The CPA laser phenomena are mediated by the excitation of the surface modes localised at the gain-loss boundaries of *PT*-symmetric structure. The Wolf-Bragg resonances of Bloch waves in the whole stack of the layers create additional modulation of $|\lambda_{1,2}|$ and frequency shift of singular points. The frequencies of transition between symmetric and broken phases significantly vary with thicknesses of an individual layers of the stack, moving to higher frequency for thinner layers. For increasing angles of incidence and number $N$ of stacked unit cells, the transition in periodic structures occurs at lower frequencies. In some particular cases the frequency phase transition will be suppressed by Wolf-Bragg resonances.

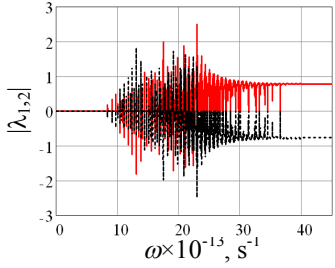

Fig. 2. Modulus of eigenvalues of $S$-matrix for 1D $PT$-symmetric periodic stack as a function of frequency at $\varepsilon'=2.1$, $\varepsilon''=0.1$, $\Theta_i=10°$, $d_m$=100 μm, $d_c$=0 mm, $N$=10. Solid red curves correspond to the eigenvalues $\lambda_1$, dashed black curves are for $\lambda_2$.

The numerical examination of the eigenvalue dependence for $PT$-symmetric stack with a passive region shows that internal resonances in the real-index region inserted between two $PT$-symmetric periodic structures causes the effective $T$-breaking perturbation strength to oscillate with frequency. As a result, the whole system reenters the symmetric phase periodically [22]. It is necessary to note that the $S$-matrix transition points move toward lower frequencies for increasing thicknesses $d_c$.

B. *Spectral Efficiency of the Combinatorial Frequency Generation*

To gain insight in the properties of CFG, it is necessary first to examine the stack linear reflectance $R(\omega)$ which is responsible for the pump wave amplitudes inside the layers and, respectively, coefficients $N_{1,2}^{\pm}$ for the fields in (1). The linear reflectance $R(\omega)$ depends on the incidence angle and is significantly influenced by the stack composition. To illustrate the latter effect, the reflectivity $|R^{(L,R)}|^2$ and transmittivity $|T|^2$ of TM waves incident at slant angles $\Theta_i=10°$ on the $PT$-symmetric stack without an embedded anisotropic dielectric layer ($d_c$=0) is displayed in Fig.3a,b. The flux-conserving anisotropic transmission resonance (ATR) phenomena, which feature zero reflectance only for incidence from one side of the structure, can be observed for frequency ranges corresponding to the $PT$-symmetric (Fig.3a) and broken-symmetry (Fig.3b) states. Some zeros of reflectance and corresponding ATRs are marked by vertical dotted gray lines. For the chosen parameters of the layers at low frequencies before symmetry breaking points ATRs can occur for both left and right incidence. At the broken-symmetry frequency range we observe only ATRs for right side incidence. The earlier studies have demonstrated that the ATR frequencies do not depend on the number of structure periods and are determined only by the parameters of the layers in the unit cell and the incidence angle [23]. All other reflectionless resonances are Wolf-Bragg resonances which do not depend on direction of incidence. The amplification of reflected waves is connected with a constructive interaction between the forward- and backward-propagating waves. Let us note that for $|T|^2>1$ the conservation relation should be written as $|T(\omega)|^2-|R^{(L)}(\omega)R^{(R)}(\omega)|=1$ [21]. Peaks of reflectivity and transmittivity after $PT$-breaking transition point correspond to the CPA laser resonant points. The example of CPA-lasing resonances is marked by vertical dotted green line (see Fig.3b). The total number of such resonances significantly vary with number of unit cells. This effect is attributed to the additional modulation of the $S$-matrix eigenvalues at Wolf-Bragg resonances. The resonant phenomena do not exist in the high frequency regions. This effect is directly related to the fact that in the high frequency limit $|R^{(L)}(\omega)R^{(R)}(\omega)|\to 1$, where $|R^{(L)}(\omega)|^2>1$ and $|R^{(R)}(\omega)|^2<1$, and correspondingly $|T(\omega)|^2\to 0$.

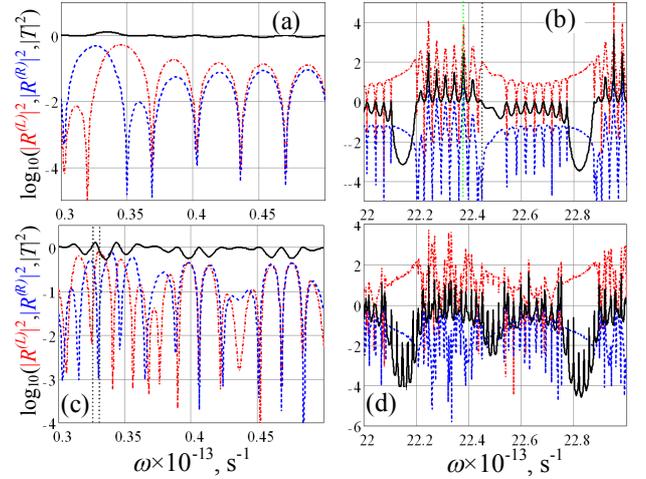

Fig. 3. TM wave reflectance ($|R^{(L)}|^2$ – dash-dot red curve, $|R^{(R)}|^2$ – dashed blue curve) and transmittance (solid black curve) for $PT$-symmetric periodic stacks with $N$=5, (a,b) $d_c$=0 and (c,d) $d_c$=2 mm, illuminated at the incidence angle $\Theta_i$ =10°.

Examination of Figs.3c (symmetric phases) and 3d (broken phases) for periodic stacks with an embedded anisotropic dielectric layer shows the significant changes in the reflectance and transmittance of electromagnetic waves. Indeed, the presence of "defect" real-index layer in the $PT$-symmetric periodic composition entails additional ATRs connected with internal resonances in the anisotropic dielectric layer and numerous CPA-lasing resonances caused by the modulation at Wolf-Bragg resonances of full stack. These differences in the magnitudes of the refracted pump waves inside the stack affect the frequency mixing efficiency as discussed below.

To assess the effect of the pump wave reflectance on CFG, intensities $|F_{r,t}(\omega_3)|$ of the waves of combinatorial frequency $\omega_3$ emitted from the *PT*-symmetric periodic stacks with an embedded nonlinear anisotropic dielectric layer have been calculated at fixed frequencies $\omega_2$ of a pump wave incident at $\Theta_{i2} = 0°$ and swept frequency $\omega_1$ of the other pump wave incident at $\Theta_{i1} = 10°$. Comparison of Fig. 4 and Figs. 3c,d demonstrates definite correlation between $|F_{r,t}|$ and $|R(\omega)|^2$. Indeed, for the interaction of waves with frequencies below the phase transition frequency (Fig.4a) $|F_{r,t}|$ reach their peaks at the frequencies corresponding to the minima of the pump wave reflectivity, thus confirming that the pump wave refraction into the stack has significant effect on the frequency mixing efficiency. Fig. 4a also shows that the peak intensity of $|F_{r,t}|$ grows with $\omega_1$. This effect can be attributed to the increase of the pump wave interaction length at the higher frequencies. For the interacting frequencies beyond phase transition point the intensities of generated waves can exhibit giant growth if at least one frequency is close to CPA-lasing state (Fig.4b). Examination of the frequency dependencies for $|F_{r,t}|$ in Fig. 4b shows that when $\omega_2$ is close to CPA-lasing point and $R^{(L)}(\omega_2) > 1$, $R^{(R)}(\omega_2) \approx 1$ and $T(\omega_2) > 1$, the fields radiated at frequency $\omega_3$ in reverse direction are amplified and $|F_r|$ has maxima at $\omega_1$ corresponding to the ATR and Wolf-Bragg resonances, At the same time, the emission in the forward direction is suppressed and $|F_t|$ becomes nearly 11-12 orders of magnitude smaller than $|F_r|$. It is attributed to the changes in the phase coherence of the generated combinatorial frequency products due to their internal reflection and refraction at the nonlinear layer interfaces. This effect will be additionally discussed in the next section. However, at the lower frequencies $\omega_1$ (below the phase transition frequency), the peak values of $|F_r|$ exceeds $|F_t|$ for nearly an order of magnitude.

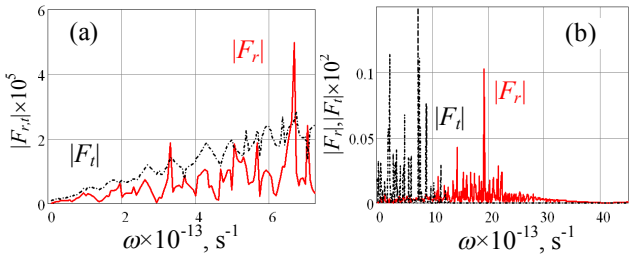

Fig. 4. Intensities $|F_{r,t}|$ of the field radiated at frequency $\omega_3 = \omega_1 + \omega_2$ in the reverse (solid red lines) and forward (dash-dot black lines) directions of the *z*-axis from *PT*-symmetric periodic stack (*N*=5) with an embedded nonlinear dielectric layer at (a) $\omega_2 = 0.35 \times 10^{13}$ $s^{-1}$ and (b) $\omega_2 = 22.76 \times 10^{13}$ $s^{-1}$. The stack is illuminated by pump waves incident at $\Theta_{i1} = 10°$ and $\Theta_{i2} = 0°$.

The enhanced CFG in the resonant structures based on *PT*-symmetric periodic structures can be attributed to the stack composition which facilitates the phase synchronism and local field intensification in the nonlinear layer due to the refraction of the pump waves by the walls of resonator. Figure 5 displays $|F_{r,t}|$ for the optical resonator with the same parameters as in Fig. 4 but inverse order of the layers in the right wall (Fig.5a). For the numerical simulations presented in Fig.5, the pump frequencies $\omega_1$ and $\omega_2$ were chosen so that frequencies of all interacting waves be below (Fig.5b) and above (Fig.5c) the phase transition point. It is necessary to note that all frequency dependences of reflectivity/transmittivity for such stack configuration are totally correlated with similar dependencies for the *PT*-symmetric walls of the resonator. Comparison of the $|F_{r,t}|$ for these two structures shows that the overall peak efficiency of CFG in the last configuration of the optical resonator is higher than in periodic stacks with embedded nonlinear dielectric layer. This implies that the efficiency of the CFG can be further increased and optimised by choosing a proper combination of the incidence angles, layer parameters and the stack composition.

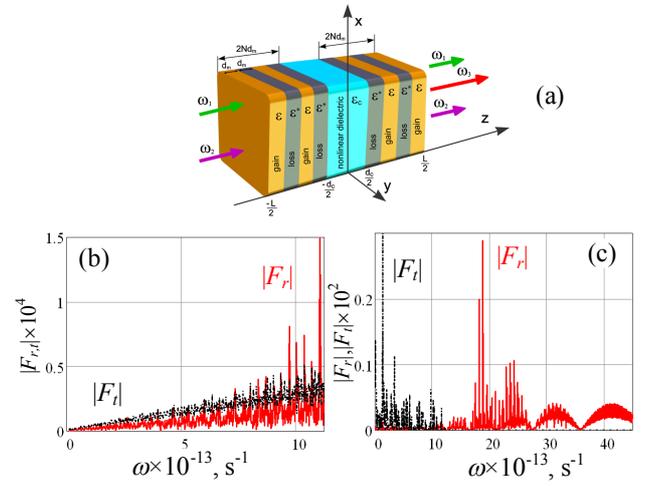

Fig. 5. (a) - Geometry for optical resonator formed by 2 *PT*-symmetric periodic structures. (b,c) - Intensities $|F_{r,t}|$ of the field radiated at frequency $\omega_3 = \omega_1 + \omega_2$ in the reverse (solid red lines) and forward (dash-dot black lines) directions of the *z*-axis from resonant structure with nonlinear cavity built up by *PT*-symmetric periodic structures (*N*=5) at (b) $\omega_2 = 0.35 \times 10^{13}$ $s^{-1}$ and (c) $\omega_2 = 22.76 \times 10^{13}$ $s^{-1}$. The stack is illuminated by pump waves incident at $\Theta_{i1} = 10°$ and $\Theta_{i2} = 0°$.

*C. Effects of Cavity and Wall Thicknesses and Resonance Enhancement of Frequency Conversion*

The intensity of the field emitted from the resonators with *PT*-symmetric structures at combinatorial frequency $\omega_3$ varies significantly with the thickness of nonlinear layer $d_c$ and thickness of *PT*-symmetric stacks $L_{PT}$. At Wolf-Bragg resonances it may reach the extremes which manifest themselves in either total suppression or giant growth of the emitted field. Based upon the analytical estimation of the $|F_{r,t}|^2$ nulls and maxima in the Section 2, the $|F_{r,t}|^2$ dependencies on $d_c$ for different number of unit cells in *PT*-symmetric walls of optical resonator in Fig.5a have been simulated and are shown in Fig. 6 for the different sets of pump wave frequencies.

At the pump frequencies corresponding to the broken-symmetry phases (Fig.6a for *N*=1) the first peak values of $|F_{r,t}|$ at $L = 1.288$ mm correspond to Wolf-Bragg resonance

($k_z^{(3)} d_c = q\pi$) of very high order ($q = 1473$). It is also noteworthy that the peak value of $|F_r|$ exceeds maximal value of $|F_t|$ for nearly five orders of magnitude. The properties of the giant Wolf-Bragg resonances at frequency $\omega_3$ depend on type of pumping wave transmittance/reflectance by the *PT*-symmetric walls. It is observed, that if both pumping waves are transmitted by the left wall without amplification ($|T|^2 \leq 1$), the peak magnitudes and variations about the mean level are reduced for higher order resonance peaks (Fig.6a). If at least one pumping wave is amplified by the wall, the enhancement of peak magnitudes and increase of variations about mean level take place (Fig.6b for $N=5$). As shown in Fig.6b, in this case $|F_r|$ exceeds $|F_t|$ for nearly 12 orders of magnitude. This result shows that the amplification of reflected combinatorial frequency wave is accompanied by huge attenuation of transmitted wave at frequency $\omega_3$. In contrast to the previous case, $|F_t|$ and $|F_r|$ are of the same order of magnitude for three interacting waves with frequencies corresponding to the *PT*-symmetry phases (Fig.6c for $N=5$). Comparison of last result with respective results for similar anisotropic nonlinear dielectric slab without *PT*-symmetric walls shows that peak intensities of generated waves for optical resonators are about one-two orders of magnitude higher.

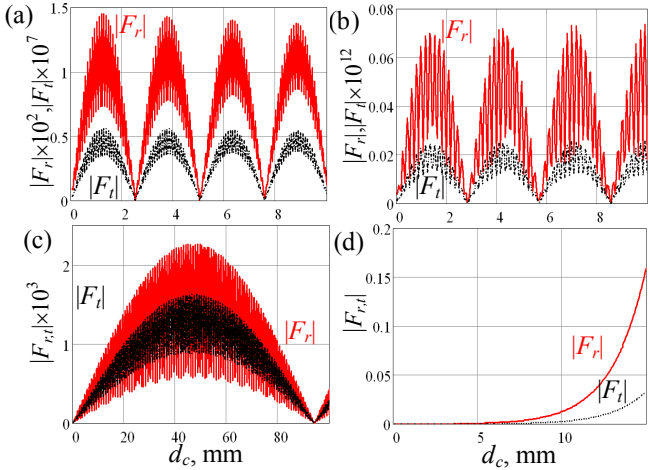

Fig. 6. Intensities $|F_{r,t}|$ of the field at frequency $\omega_3 = \omega_1 + \omega_2$ radiated from the resonator with nonlinear dielectric layer of thickness $d_c$ in the reverse ($|F_r|^2$ - solid red line) and forward ($|F_t|^2$ – dash-dot black line) directions of the *z*-axis at $\Theta_{i1} = 0°$, $\Theta_{i2} = 10°$, tg $\Delta_{xx,zz} = 0$, and
(a) $\omega_1 = 22.81 \times 10^{13}$ $s^{-1}$, $\omega_2 = 23.68 \times 10^{13}$ $s^{-1}$; $N=1$;
(b) $\omega_1 = 22.76 \times 10^{13}$ $s^{-1}$, $\omega_2 = 28.87 \times 10^{13}$ $s^{-1}$, $N=5$;
(c) $\omega_1 = 0.32 \times 10^{13}$ $s^{-1}$, $\omega_2 = 10.77 \times 10^{13}$ $s^{-1}$, $N=5$;
(d) $\omega_1 = 0.32 \times 10^{13}$ $s^{-1}$, $\omega_2 = 10.77 \times 10^{13}$ $s^{-1}$, $N=5$, tg $\Delta_{xx,zz} = 0.001$.

It is necessary to note that the number $N$ of stacked unit cells has strong impact on the efficiency of harmonic generation (see eq.(5)) and peaks of $|F_{r,t}|$ change the values. But the dependence of field intensity $|F_{r,t}|$ on the number of unit cells in the walls of resonator is nonmonotonic and is attributed to unequal variations of the reflection coefficients of the incident pump waves which cause the pump wave amplitude and phase disbalance in the mixing process.

To estimate the effect of dissipation on $|F_{r,t}|$ the structure with the same parameters as in Fig. 6a-c has been simulated in the case of imperfect nonlinear layer with the loss tangent tg $\Delta_{xx,zz} = 0.001$. The numerical simulation results show that dissipation leads to strong enhancement of $|F_{r,t}|$ at the higher orders of Wolf-Bragg resonances connecting with disbalance in the three-wavemixing process in the instable system which further increases the efficiency of the combinatorial frequency generation and suppresses the undulation of field intensity, as seen in Fig. 6d.

## IV. CONCLUSION

Combinatorial frequency generation by nonlinear resonant structures with *PT*-symmetric periodic stacks has been studied in the three-wave mixing process. The closed form solution of the nonlinear scattering problem, illuminated by a pair of obliquely incident pump waves of frequencies $\omega_1$ and $\omega_2$, has been obtained in the approximation of weak nonlinearity using the modified TMM method. The developed theory was illustrated through examples of numerical simulations and the features of CFG by *PT*-symmetric resonant structures are discussed.

The effect of the structure geometry on the spontaneous symmetry breaking in linear *PT*-symmetric resonators is studied. The performed parametric study has shown that in contrast to "ideal" *PT*-symmetric periodic layered structures, the presence of resonant cavity can substantially increase the number of ATR and CPA-lasing points. Depending on the pump wave frequencies, the $|F_{r,t}|$ properties may qualitatively alter. The simulation results have demonstrated that for the interaction of waves with frequencies below the phase transition, the nonlinear scattering coefficients $|F_{r,t}(\omega_3)|$ at the combinatorial frequency $\omega_3 = \omega_1 + \omega_2$ are strongly correlated with minima of linear reflectance $|R(\omega)|$ of the respective stacks. It is shown that the conversion efficiency can be significantly increased as one pump wave frequency is near the CPA-lasing resonances. At the same time, it has been observed that the stack composition can strongly affect the CFG efficiency, e.g. the $|F_{r,t}(\omega_3)|$ peak magnitude depend on the sequence of the layers in *PT*-symmetric periodic walls of optical resonator. This effect is caused by the local field intensification in the nonlinear layer due to the scattering of interacting waves by the *PT*-symmetric walls of the resonator.

It is shown that for special combinations of the pump wave frequencies the intensity of the scattered waves of the frequency $\omega_3$ can be dramatically intensified at the high order Wolf–Bragg resonances which satisfy the additional conditions of the spatial synchronism. The conditions for attaining the global maxima and nulls of the scattered fields have been obtained. The effect of losses in the nonlinear layer have been assessed and it is shown that they dramatically increase the intensity of the wave emitted in the reverse direction and suppress the one passing through the stack.

## References

[1] W. Cai and V. Shalaev, *Optical Metamaterials: Fundamentals and Applications*, Springer, New York, 2009.
[2] C. Denz, S. Flach, Y. S. Kivshar, *Nonlinearities in Periodic Structures and Metamaterials*, Springer, Berlin, 2010.
[3] O.V. Shramkova, A.G. Schuchinsky, "Harmonic generation and wave mixing in nonlinear metamaterials and photonic crystals", *Int. J. of RF and Microwave Computer Aided Engineering*, vol. 22, pp. 469-482, 2012.
[4] S. Zhu, Y. Zhu, N. Ming, "Quasi–phase-matched third-harmonic generation in a quasi-periodic optical superlattice," *Science*, vol. 278, pp. 843-846, 1997.
[5] Y. B. Chen, C. Zhang, Y.Y. Zhu, S.N. Zhu, H.T. Wang, N.B. Ming, "Optical harmonic generation in a quasi-phase-matched three-component Fibonacci superlattice LiTaO3," *Appl.Phys.Lett.*, vol. 78. pp. 577-579, 2001.
[6] B. M. Bertolotti, "Wave interactions in photonic band structures: an overview," *J. Opt. A: Pure Appl. Opt.*, vol. 8, S9, 2006.
[7] M.-L. Ren, Z.-Yu. Li, "Exact iterative solution of second harmonic generation in quasi-phase-matched structures," *Optics Express*, vol. 18, pp. 7288-7299, 2010.
[8] T. Paul, C. Rockstuhl, F. Lederer, "A numerical approach for analyzing higher harmonic generation in multilayer nanostructures," *J. Opt. Soc. Am. B*, vol. 27, pp. 1118-1130, 2010.
[9] J. Martorell, Broadband efficient nonlinear difference generation in a counterpropagating configuration, *Appl. Phys. Lett.*, vol. 86, p. 241113, 2005.
[10] O.V. Shramkova, A.G. Schuchinsky, "Nonlinear scattering by anisotropic dielectric periodic structures," *Advances in OptoElectronics*, p. 154847, 2012.
[11] O. Shramkova, A. Schuchinsky, "Combinatorial frequency generation in quasi-periodic stacks of nonlinear dielectric layers," *Crystals*, vol 4, pp. 209-227, 2014
[12] C. M. Bender and S. Boettcher, "Real Spectra in Non-Hermitian Hamiltonians having *PT* symmetry," *Phys. Rev.Lett.*, vol. 80, pp. 5243-5246, 1998.
[13] Z. Lin, *et al.*, "Unidirectional invisibility induced by *PT*-symmetric periodic structures," *Phys. Rev. Lett.*, vol.106, p. 213901, 2011.
[14] Yo. Sun, *et al.*, "Experimental demonstration of a coherent perfect absorber with *PT* phase transition," *Phys. Rev. Lett.*, vol. 112, p. 143903, 2014.
[15] K. G. Makris, *et al.*, "Beam dynamics in *PT* symmetric optical lattices," *Phys. Rev. Lett.*, vol. 100, p.103904, 2008.
[16] M. C. Zheng, *et al.*, "*PT* optical lattices and universality in beam dynamics," *Phys. Rev. A*, vol. 82, p. 010103, 2010.
[17] N. Lazarides and G. P. Tsironis, "Gain-driven discrete breathers in *PT*-symmetric nonlinear metamaterials," *Phys. Rev. Lett.*, vol.110, p. 053901, 2013.
[18] S. Longhi, "*PT*-symmetric laser absorber," *Phys. Rev. A*, vol 82, p. 031801(R), 2010.
[19] N. Blombergen, *Nonlinear Optics: A lecture note*, Benjamin, New York-Amsterdam, 1965.
[20] A. Yariv and P. Yeh, *Waves in Crystals: Propagation and Control of Laser Radiation*, Wiley, New York, 1984.
[21] Li Ge, Y. D. Chong, and A. D. Stone, "Conservation relations and anisotropic transmission resonances in one-dimensional *PT*-symmetric photonic heterostructures," *Phys. Rev. A*, vol. 85, p. 023802, 2012.
[22] Y.D.Chong, Li Ge A.D. Stone, "Coherent perfect absorbers: time-reversed lasers," Phys. Rev. Lett., vol.106, p. 093902, 2011.
[23] O.V. Shramkova, G.P. Tsironis, "Wave scattering by PT-symmetric epsilon-near-zero periodic structures," *European Microwave Conference (EuMC 2015)*, Paris, France, September 6-11, 2015.

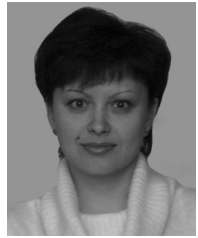

**Oksana V. Shramkova** (M'07-SM'08) received M.Sc.degree in Solid-State Physics from the National Technical University ''Kharkov Polytechnic Institute'' (NTU-KhPI) in 1998 and PhD degree in Radiophysics from the Usikov Institute of Radiophysics and Electronics of the NAS of Ukraine (IRE NASU) in 2001.

From 2001 to 2010, she was with the Department of Solid-State Radiophysics of IRE NASU, Ukraine. From 2010 to 2014, she was a Marie Curie Research Fellow in the School of Electronics, Electrical Engineering and Computer Science at the Queen's University of Belfast. Since 2014 she is a Senior Researcher at the Center for Quantum Complexity and Nanotechnology of the Physics Department of the University of Crete. She has authored and co-authored about 120 papers in referred Journal and Conference Proceedings. Her research interests include physics-based modeling of linear and nonlinear phenomena in complex electromagnetic structures and metamaterials.

In 2006, Dr. Shramkova was awarded the academic title of Senior Research Scientist. She is also a Senior member of IEEE. In 2008, she was awarded State Prize of the President of Ukraine for young researchers (top prize for young researchers).

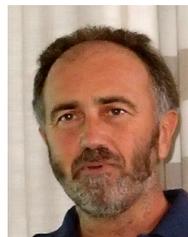

**Giorgos P. Tsironis** obtained his Ph.D. degree in theoretical condensed matter and statistical physics from the University of Rochester in 1987.

From 1987 to 1996, he was a Postdoctoral Associate in the University of California San Diego and in the Fermi National Accelerator Lab, and assistant professor of Physics at the University of North Texas. He joined the Department of Physics of the University of Crete in 1994. From 2007 to 2011, he served as Chairman of the Department of Physics. Currently, he is Professor of Physics at the Physics Department of the University of Crete and coordinator of the Crete Center for Quantum Complexity and Nanotechnology at the Physics department of the University of Crete, also he leads the Nonlinear and Statistical Physics Group at the IESL-FORTH. He has published over 150 papers in refereed journals. His research interests are in the areas of condensed matter physics, statistical mechanics, nonlinear physics, and quantum metamaterials. Important contributions relate to the nonequilibrium statistical mechanics of discrete breathers, nonlinear and superconducting metamaterials.